# Revealing the Weaknesses of File Sharing System on Cloud Storages

Fauzi Adi Rafrastara        Qi DeYu

*Abstract*—Cloud storage provides the simpler way to share the files privately and publicly. A good Cloud Storage Provider (SCP) is not only measured by the access speed or file size that can be shared to others, but also regarding the security issues in file sharing itself. In this paper, we analyze the security of file sharing in 3 Chinese CSPs, which are: Baidu, Weiyun and Kanbox. Those CSPs have their own vulnerabilities that successfully revealed. We also provide some suggestions to countermeasure the weaknesses so that they can maintain the quality while improving the security.

## I. Introduction

Inside the cloud computing system, there is a cloud service that earlier it was very famous to be classified as: Software as a Service (SaaS), Platform as a Service (PaaS) dan Infrastructure as a Service (IaaS). Along with the advance technology and invention, currently there is no category on cloud service. Those now are in one category, called Everything-as-a-Service (XaaS) [1]. It is because most of the CSPs (Cloud Service Provider) now combine those 3 services or sometime plus some other services, such as: Storage as a Service, Network as a Service or even Monitoring as a Service, and then provide it to users as one bundled system.

One of cloud computing products that currently booming in many parts of the world, is Cloud Storage. It has been experiencing the tremendous growth in the past 5 years. In China, there is a kinds of war when many giant IT companies suddenly create a new product and become Cloud Storage Provider (CSP), and trying to get as many as possible users by promoting their product massively with offering some extraordinary features [2].

When the western CSP are offering free capacity to the new users around 1 to 2 digits of Giga Bytes, the Chinese CSPs are currently offering free capacity around 1 to 2 digits of Tera Bytes. At least there are 4 CSPs that have such kinds of special promotion. However, size of capacity is not the only consideration that users should think firstly. The security level becomes urgent when the cloud storage is used to save the private data.

Fauzi Adi Rafrastara
South China University of Technology
China

Qi Deyu
South China University of Technology
China

This paper consists of 6 sections. The 2nd section will discuss about cloud storage. The 3rd section will give the discussion in more details about sharing methods in CSP. The security in CSP, especially in term of file sharing, will be explained thoroughly in the 4th section. Assessment and Suggestion will be in the 5th section and followed by the Conclusion as a last one.

## II. Cloud Storage

The existence of cloud storage is very important nowadays. The need for this service is getting bigger and bigger, as they provide the easiness for the people to access, synchronize, share and backup data easily. Users will be able to access their digital content any time, from anywhere, and with any device (smartphone, tablet, notebook, or desktop PC) [3].

Since some western CSPs were blocked in China, some local CSPs raised with fantastic offer, such as: Baidu (up to 2TB storage capacity), Alibaba Kanbox (up to 10TB) and Tencent Weiyun (up to 10TB) for free. Such offer is tempting so many people especially in China and it is successfully make them ignore the dropbox, googledrive, etc. But actually, no matter how large the capacity offered, the security issue cannot be denied that it is still very important things to be considered. As mentioned by Zhou et al [4] in their paper, security and privacy issue is regarded as a top concern among 9 challenges in cloud computing system.

Cryptographic mechanism is used to secure all communication between users and CSPs, such as uploading and downloading the data [5] [3]. Another security aspect that should be given more attention is the security of file sharing [3]. It is not a small thing that will not bring a serious problem to the users. When we begin to share the file, it means that we are ready to take all risk for that. The fact is, by sharing the files with some other users, sometimes it opens security hole. The afterwards discussion will explain the situation by analyzing 3 popular Chinese CSPs, those are: baidu [6], weiyun [7] and kanbox [8].

## III. Sharing Methods on Cloud Storage

The importance of sharing features in cloud storage is to make the sharing file becomes much easier without using email attachments which sometimes have limitation in term of file size [9]. On the other hand, sharing methods in cloud storage also have some functionality which does not exist in sharing through email.

According to Chu et al [3], there are 3 kinds of sharing methods in cloud storage (see table I). Here we will discuss it one by one and get to know the sharing methods provided by 3 CSPs which being focused on this research.

TABLE I.     FILE SHARING METHODS USED BY BAIDU, WEIYUN AND KANBOX

| Types | Baidu | Weiyun | Kanbox |
|---|---|---|---|
| Public Sharing | ✓ | ✗ | ✓ |
| Private Sharing | ✓ | ✓ | ✗ |
| Secret URL Sharing | ✓ | ✗ | ✓ |

### A. *Public Sharing*

There's no access control on this kind of sharing method. The data is intended for the public, so anyone can get the data without any authentication or authorization. In this scenario, the data owner can generate the URL of the document and publish it on the website. Afterwards anyone on the internet can access or even download this document directly through the given URL. In another scenario, the contents that shared publicly on some CSPs can be accessed through Google search as well.

Weiyun does not have the public sharing option. Anyone who wants to access the documents stored inside the Weiyun Cloud Storage, they have to sign-in first, except for the image file. Here, anyone can still see the image with the medium resolution without sign-in to CSP. But they are required to have the account when they want to access the image with real resolution. On the contrary, both baidu and kanbox have this option.

### B. *Private Sharing*

Authentication is required here. Firstly, the owner must specify who will get the access for the data. Afterwards the CSP will authenticate anyone which trying to access the data whether they are in the list or not. They must sign-in first, and their identity usually will be shown on the owner's CSP window.

Among the 3 Chinese CSPs, only Kanbox that does not have private sharing option. It only provides sharing method through public URL and via e-mail. On the other hand, private sharing method is provided in Baidu and Weiyun.

### C. *Secret-URL Sharing*

Secret-URL Sharing can be a bridge between public and private sharing. The URL of public sharing is not distributed in a secret way. While private sharing, it is a secret way but need the account to access it. By being the middle option, Secret-URL Sharing provides the secret distribution with the open access account.

In this method, the URL of documents that is going to be shared will be distributed secretly through private way, such as e-mail. So, only the one who received the e-mail from data's owner can access the contents without further authentication or authorization. This option is available in Baidu and Kanbox.

Weiyun actually does have the option to share the URL through e-mail. However, it cannot be categorized as a Secret-URL Sharing since anyone who receives the email remains required to sign-in with QQ or Weiyun account.

### D. *File Sharing Security*

File sharing in cloud storage is one of the most interesting topics to be discussed, in which the mechanism of this method is different compared to file sharing via email attachment. By using e-mail attachment, the shared data from Allice will permanently exist in Bob's e-mail storage. But in cloud storage, Allice can decide when she will open and close the door to share the documents anytime. It is depend on the features provided by CSPs.

According to Chu et al. (2013) [3], there are 8 parameters that can be used to assess the security of CSPs in term of data sharing. By using those 8 plus 1 additional parameter, this paper is going to reveal the security hole of 3 Chinese CSPs, those are: Baidu, Weiyun and Kanbox.

### E. *Non-Dead URL*

This parameter refers to the link that remains working when the file has been updated, deleted or even replaced by a new one with the same name.

In this case, Weiyun apparently does not change the URL when the file has been updated. Suppose that Allice wan to share a document with Bob by using secret-URL sharing. Bob will do have the access to her file permanently as long as that file is existing. If they both are using Weiyun, then Bob will be able to access the file even though it has been updated many times by Allice. When Allice deletes the file, then it will be ended. The URL on the Bob's pocket will be deactivated.

It is probably good news when Allice works together with Bob and she shares the document through Weiyun. Bob will be able to monitor the progress of Allice because he can get the up-to-date version of the document with one single link only. But in another scenario, it could be dangerous when Allice does not have intention to share the documents permanently. She might do not realize with the situation in which Bob can easily collect a lot of information without her permission, by monitoring the update on Allice's document.

TABLE II.     NONDEAD URL

| URL still active when: | Baidu | Weiyun | Kanbox |
|---|---|---|---|
| Update | ✗ | ✓ | ✗ |
| Delete | ✗ | ✗ | ✗ |
| Replace (creating a new file with the same name) | ✗ | ✗ | ✗ |

On the other hand, Baidu and Kanbox have the more secure way about it. The URL is not a "NonDead" link since they will change the URL when there is an update on the file. (See Table II).

### F. *Uncertain Identities*

The objective of private sharing is to make sure that only the authorized people can access the data. But in fact, not all CSPs follow this rule. The owner should be able to see the identity of the users who access the data. But sometimes, there are some security holes so that the owner cannot identify it easily as explained by Chu et al in their paper [3].

Let say Allice shares the URL to Bob and John by using private sharing. Since Bob has 2 accounts in this CSP, then Bob tries to sign-in with his second account who has not listed in Allice data yet. After Bob has successfully signed-in, Allice will be curious with that strange account. If only 2 accounts that the file is shared with, it could be easy to confirm whether this strange account belongs to Bob or John. But, what if we share the document to more than 50 or hundred peoples? It would be difficult to clarify once we found the uncertain identities. Among 3 CSPs, it is only Kanbox who does not have the private sharing methods. So Kanbox will be ignored in this part.

Baidu provides the private sharing method. Here, data owner can share the document to the colleagues that have been registered and listed as "good friends" (好友). And anyone who has listed as a "good friends", it is confirmed that they already have the Baidu's account before. Once Allice wants to share the document with others, she only needs to pick it from the "good friends" list and the notification will be sent privately to the expected users. Suppose that Bob has 2 Baidu's accounts, and Allice has sent the notification to his first account, then he can only access the data from his first account. Here, the shared data will only appear on the selected users account (required to sign-in), and there is no specific URL appears when the shared data is being accessed. The peoples who have the privilege to access the data will be shown on the Data Sharing Window with some modifications to hide and protect the full identity.

*Example 1:*

"*username@email.com*", *will be shown as*
"*us…e@email.com*"

Discussing about Weiyun, actually the user interface of Weiyun CSP is very simple. There are only few features included. Especially on sharing mechanism, it provides 2 options only, those are: by url" and "by mail". There is no specific option to do private sharing. However, it is compulsory to sign-in first using QQ or Weiyun account to access the shared file. This condition is suitable with the requirement of private sharing where the users need to have an account access. By this reason, Weiyun is considered have the private sharing method.

Consider that Allice wants to share the document with Bob by choosing "by mail" option, and then Bob click the link and sign-in to the Weiyun CSP, it still difficult to identify who are accessing that document. There is no list of peoples who have privilege to the document. Allice is unable to check whether Bob has accessed it or not. Even Allice is also unable to maksure that Bob accessed the file using his own account or not. And it is also impossible for Allice to know whether there is any other person or not accessing her shared file.

### G. *Unauthorized Resharing*

Unauthorized Ressharing is the condition where Allice has shared a link to Bob privately, but Bob turns out re-shares again to others, and it is still possible for others to re-share again many times. Allice cannot control the link once she releases the URL, both publicly and privately.

This parti is also for the CSPs that have private sharing feature. Once again, Kanbox will be ingnored.

On Baidu, there is a private way to share the data. No one can open the door if they are not invited by the owner. Bob can receive the shared file from Allice but no way to do direct re-sharing. If Bob wants to do such action, he has to keep the file first to his cloud storage then share it manually. But it is a different scenario.

Weiyun differs from Baidu in term of this parameter. There is no private way provided by Weiyun. Anyone can access the link given by the owner as long as they sign-in to Weiyun Cloud Storage account. The owner can send the link to some expected colleagues. But in the same time, the owner cannot control if one of those colleagues is trying to re-share the given link by spreading from their account. This condition allows for what so called "Unauthorized Resharing".

### H. *Indiscriminate Accessing URL*

It is happened when there is no difference between URL used by Allice (as an owner) and Bob to access the file. On their paper, Chu et al [3] mentioned that Google Drive have such kind of weakness. They illustrate that if Alice is in a meeting and accesses her file while using a projector, the URL will be shown in the browser's address bar. It is dangerous since in Google Drive, any file is accessed via the associated URL, even for the file owner.

Unlike a Google Drive, these 3 Chinese CSPs are showing the page's URL, not file's URL on the address bar of web browser. This kind of URL is relatively safer and inaccessible by the unauthorized user.

### I. *Non-HTTPS URL for Sharing*

HTTPS is important to secure the communication line in internet era. It protects not only the transmission of the involved data but also the resource locator [3]. HTTPS actually is combination of HTTP and SSL/TLS to secure the communication between Web browser and Web server, especially to prevent the kinds of web attacks, including eavesdropping [10] [11].

Unfortunately, these 3 Chinese CSPs are not using the HTTPS as their communication protocol, so that the integrity and confidentiality of the data are at stake.

## J. Non-HTTPS Shortened URL

URL Shortening is a technique to decrease significantly the length of the URL so that the URL becomes easy to read, remember and share [12]. When the website is accessed from the mobile phone, then the user will very easy to copy-paste, share or even to spell the URL.

There is only one CSP among 3 Chinese CSPs being studied in this paper that have URL Shortening feature, called Weiyun. However, as mentioned in the previous discussion, Weiyun does not use the secure protocol for the communication between client and server.

## K. No Privacy on Sharing

On private sharing, this term means anyone who are invited to enjoy the shared file can also see other people identities with no filter, i.e. full email address as happened in Dropbox and Google Drive [3]. It is dangerous if other users do not have good intentions to exploit this weakness for the sake of them. Since the problem relates to private sharing, then Kanbox will be ignored on this discussion.

Talking about Baidu and Weiyun, both of them cannot be considered that they are running "No Privacy on Sharing". Baidu does not have this weakness. But, it does not mean that Baidu has a very powerfull security. This is simply because Baidu do not have the feature to show the people who connected to the shared files. While on the Weiyun, the invited guests will be listed neatly and safely. It is only email address of the inviter that will be shown completely. Other email address will be secured by hiding some characters of account name (see Example 1).

## L. Sharing of Trash Files

A serious problem happened in Google Drive when the people still can access the file that has been deleted and store it into trash folder [3]. Alice deletes the file probably because she does not want to share it anymore. But unfortunately she might be not aware that people can still access it even in the trash.

After we analyzed the 3 Chinese CSPs, we conclude that all of them are free from this problem. Baidu and Kanbox will change the URL every time when changes happened. They automatically deactivate the previous link that has been shared before. On the other hand, Weiyun still maintain the same link if owner only update the contents. But when owner deletes the file, then the link will be removed automatically.

## M. Fixed URL

The scenario of this parameter according to Chu et al [3] is as follows:

Suppose that Allice has shared the URL publicly, but then suddenly she changes her mind and she wants to share it privately, so what happened with the previous URL?

Chu et al [3] reported that Google Drive uses the same URL for that scenario. It is unsecure since now Allice wants to categorize the file as a private, but in fact the URL has been widely spread as public URL. If earlier Bob already got the link when Allice made it open for public, then now Bob can still access this secret file. Allice probably does not aware with this situation.

On Baidu, for the scenario above, the file turns out still accessible using public URL. So, it is like no privilege, once the URL has been publicly shared. But there is good news that Baidu provides a feature to cancel the file sharing. If Alice cancels the public sharing in advance, then it will be secure when she wants to re-share it privately.

On the other hand, Weiyun and Kanbox both cannot be measured using this parameter, because Weiyun does not have public sharing feature, whereas Kanbox does not provide the private sharing feature.

## IV. Assessments and Suggestions

According to the prior discussion, we provide a table that can explain the vulnerabilities of 3 Chinese CSP, namely: Baidu, Weiyun and Kanbox, in a simpler way (see Table III).

TABLE III.   SECURITY ASSESSMENTS OF BAIDU, WEIYUN AND KANBOX

| Vulnerabilities | Baidu | Weiyun | Kanbox |
|---|---|---|---|
| Non-Dead URL | ✗ | ✓ | ✗ |
| Uncertain Identities | ✗ | ✓ | O |
| Unauthorized Resharing | ✗ | ✓ | O |
| Indiscriminate Accessing URL | ✗ | ✗ | ✗ |
| Non-HTTPS URL for Sharing | ✓ | ✓ | ✓ |
| Non-HTTPS Shortened URL | O | ✓ | O |
| No Privacy on Sharing | ✗ | ✗ | O |
| Sharing of Trash Files | ✗ | ✗ | ✗ |
| Fixed URL | ✗ | O | O |

The assessments and suggestions regarding the security weaknesses of 3 Chinese CSPs will be discussed as follows:

Baidu and Kanbox do the good way, whereas Weiyun should be more careful about the "Non-Dead URL" issue. So, to the best of our knowledge, the link should be invalidated when the file is changed (updated or removed). The CSPs can also provide the option when the owner wants to share the live file, so that people can access the updated file anytime without changing the URL.

It is better to implement the private sharing methods as provided by Baidu, since it can make sure that the file is only accessed by the invited people with the registered account. However, especially for the owner, it is important to consider that owner should be able to check which users that currently on-line and accessing the shared file.

Baidu has a good security system against "unauthorized re-sharing". There is no URL exposed in private sharing, and it can reduce the possibility to do unauthorized re-sharing. Other CSPs are recommended to implement such kind of private sharing method. If insisted using URL on private sharing, then they have to provide another security method to generate the URL that can be accessed once only. Once the users accept the invitation and sign-in to their account, then the URL will be deactivated immediately [3].

These 3 Chinese CSPs are using the different URL for sharing and for owner. It means that, they provide a good security against "indiscriminate accessing URL".

Up to the writing of this paper is completed, we have not yet found the reason why these Chinese CSPs are not using HTTPS as their protocol. Since cloud storage is involving client-server system and private data, it is required to secure the communication line by implementing HTTPS.

Chu et al mentioned that no shortening service supports SSL, so "secret" URLs should not be shortened [3]. Howevery, security is more important than Beauty.

When the owner deletes the file from the cloud, the link to this file should be disabled immediately [3]. These Chinese CSPs are implementing such kind of concept so they are free from this threat.

Baidu provides a proper way in private sharing, especially regarding "no privacy on sharing" issue. But it is better if owner can choose whether he or she will allow other people to see the sharing list of the shared file or not [3].

The URL should not be fixed when there is a change in sharing method. CSPs have to be careful for such unexpected vulnerability, and they have to make sure that every changing on sharing settings will be followed by the changing of URL of the shared file. Among these 3 Chinese CSPs, only Baidu that can be measured using this issue and it can handle it well. Later, when Weiyun and Kanbox are going to provide both public and private sharing, we suggest to consider such security hole. Baidu can be the good model for countermeasures this problem.

## v. Conclusion

Recently, storage media have been experiencing the tremendous growth, which are now widely available the virtual hard drive, namely Cloud Storage. On the other hand, file sharing is a kind of daily activity that cannot be denied by those who are active in using Information and Technology product.

These 2 things currently have a close relationship where cloud storage can make the sharing process become simpler. However, we should aware for all kinds of vulnerabilities that may be not previously realized.

This paper was discussing about the security of file sharing system on Cloud Storage, especially by analyzing 3 Chinese CSPs: Baidu, Weiyun and Kanbox. Depart from the basic of some parameters proposed by Chu et al [3], we did some experiments to test their security in file sharing. As a result, there were some vulnerabilities found in each CSP that can be seen in Table 3.

Baidu, Weiyun and Kanbox have their own vulnerabilities. By looking at the Table 3, at least users will know the security level of each CSP, especially in term of file sharing system. Once more, discussing the security of the system is very interesting since there is no perfect system. By this research, it is mandatory for users to be careful when they share the files in the cloud.

### References

[1] S. J. Nirmala, S. M. S. Bhanu, and A. A. Patel, "A Comparative Study of The Secret Sharing Algorithms for Secure Data in The Cloud," *International Journal on Cloud Computing: Services and Architecture (IJCCSA)*, vol. 2, no. 4, pp. 63-71, August 2012.

[2] Cindy Kuan. (2014, Augustus) Creativehunt. [Online]. http://www.creativehunt.com/shanghai/articles/dropbox-alternatives-chinese-cloud-storage-services

[3] Cheng-Kang Chu et al., "Security Concerns in Popular Cloud Storage Services," *Pervasive Computing, IEEE*, vol. 12, no. 4, pp. 50-57, October-Desember 2013.

[4] Minqi Zhou, Rong Zhang, Wei Xie, Weining Qian, and Aoying Zhou, "Security and Privacy in Cloud Computing: A Survey," in *Sixth International Conference on Semantics, Knowledge and Grids*, Beijing, 2010, pp. 105-112.

[5] Renjith P and Sabitha S., "Survey on Data Sharing and Re-Encryption in Cloud," *International Journal of Advanced Research in Computer Engineering & Technology (IJARCET)*, vol. 2, no. 2, pp. 477-480, February 2013.

[6] Baidu Yun WangPan (Baidu Cloud Storage). [Online]. http://pan.baidu.com

[7] Weiyun. [Online]. http://www.weiyun.com

[8] KuPan. [Online]. http://www.kanbox.com

[9] Technology and Research (A*STAR) The Agency for Science. (2014, June) Science Daily. [Online].
http://www.sciencedaily.com/releases/2014/06/140619145926.htm

[10] Douglas Jacobson and Joseph Idziorek, Computer Security Literacy: Staying Safe in Digital World. Boca Raton, Florida, USA: CRC Press, 2013.

[11] William Stallings, Cryptography and Network Security: Principles and Practice, Fifth Edition. Beijing, China: Publishing House of Electronics Industry, 2011.

[12] Christopher C. Elisan, Malware, Rootkits & Botnets: A Beginner's Guide. USA: McGraw Hill, 2012.